\documentclass[aps,onecolumn,12pt,showpacs,preprintnumbers]{revtex4}

\usepackage{amsmath}
\usepackage{bm}
\usepackage{graphicx}
\usepackage{epstopdf}

\begin{document}

\title{Dark Matter Axions Revisited}

\author{Luca Visinelli}
\email[Electronic address: ]{visinelli@utah.edu}
\author{Paolo Gondolo}
\email[Electronic address: ]{paolo@physics.utah.edu}
\affiliation{Department of Physics, University of Utah, 115 S 1400 E $\#$201, Salt Lake City, UT 84102, USA.}
\date{\today}

\begin{abstract}
We study for what specific values of the theoretical parameters the axion can form the totality of cold dark matter. We examine the allowed axion parameter region in the light of recent data collected by the WMAP5 mission plus baryon acoustic oscillations and supernovae, and assume an inflationary scenario and standard cosmology. We also upgrade the treatment of anharmonicities in the axion potential, which we find important in certain cases. If the Peccei-Quinn symmetry is restored after inflation, we recover the usual relation between axion mass and density, so that an axion mass $m_a =(85\pm3){\rm~\mu eV}$ makes the axion 100\% of the cold dark matter. If the Peccei-Quinn symmetry is broken during inflation, the axion can instead be 100\% of the cold dark matter for $m_a < 15{\rm~meV}$ provided  a specific value of the initial misalignment angle $\theta_i$ is chosen in correspondence to a given value of its mass $m_a$. Large values of the Peccei-Quinn symmetry breaking scale correspond to small, perhaps uncomfortably small, values of the initial misalignment angle $\theta_i$.
\end{abstract}


\maketitle

\section{Introduction} \label{introduction}
The recent measurements by the WMAP mission \cite{komatsu} have established the relative abundance of dark and baryonic matter in our Universe with great precision. About 84$\%$ of the content in the Universe is in the form of cold dark matter (CDM), whose composition is yet unknown. One of the most promising hypothetical particles proposed for solving the enigma of the dark matter nature is the axion \cite{weinberg} \cite{wilczek}. This particle was first considered in 1977 by R. Peccei and H. Quinn \cite{peccei} in their proposal to solve the strong-CP problem of the QCD theory. Although the original PQ axion is by now excluded, other axion models are still viable \cite{Kim:1979if, shifman, dine, zhitnitskii}.

The hypothesis that the axion can be the dark matter particle has been studied in various papers (see e.g. \cite{preskill, abbott, dine1, stecker, wilczek1, lyth, beltran, hertzberg} and the reviews in \cite{fox, sikivie}). Here we examine the possibility that the invisible axion may account for the totality of the observed CDM, in the light of the WMAP5 mission, baryon acoustic oscillations (BAO) and supernovae (SN) data. We also upgrade the treatment of anharmonicities in the axion potential, which we find important in certain cases. We consider invisible axion models, in which the breaking scale of the PQ symmetry $f_a$ is well above the electroweak scale. The axion parameter space is described by three parameters, the PQ energy scale $f_a$, the Hubble parameter at the end of inflation $H_I$ and the axion initial misalignment angle $\theta_i$.

\section{Axion properties}

The solution to the strong CP problem proposed by R.~Peccei and H.~Quinn \cite{peccei} introduces a new $U(1)_{PQ}$ symmetry in the theory of strong interactions. The Peccei-Quinn theory in its simplest version depends on a unique parameter, the energy scale at which the $U(1)_{PQ}$ symmetry is broken. The axion field $a(x)$ originates from the $U(1)_{PQ}$ symmetry breaking at a temperature $T \sim f_a$. Axions are the quanta of the axion field \cite{weinberg, wilczek1}.

The temperature-dependent axion mass arises through instanton effects and is given by \cite{gross}
\begin{equation} \label{axion_mass}
m_{a}(T) =
\begin{cases}
m_a b \big(\frac{\Lambda}{T}\big)^4 ,& T\gtrsim \Lambda,\\
\quad m_a, & T\lesssim \Lambda.
\end{cases}
\end{equation}
We set $\Lambda = 200$ MeV \cite{kolb}, and take the model-dependent factor $b=0.018$, consistently with previous work \cite{beltran, fox, hertzberg}. The zero-temperature mass $m_a \equiv m_a(T=0)$ is \cite{weinberg}
\begin{equation}
\label{eq:axionmass}
m_a = \frac{\sqrt{z}}{1+z}\frac{f_{\pi}m_{\pi}}{f_a/N} = 6.2 {\rm \mu eV}\bigg( \frac{10^{12}{\rm GeV}}{f_a/N}\bigg),
\end{equation}
where $z \simeq 0.56$ and $m_{\pi}$ and $f_{\pi}$ are the pion mass and decay constant respectively. The integer $N$ represents the $U(1)_{PQ}$ color anomaly index; in this paper we set $N=1$, consistently with \cite{sikivie}.

In the following, we set $Q^2 = \frac{\sqrt{z}}{1+z} f_{\pi}m_{\pi}= (78.7$ MeV)$^2$. From $m_a(T)$ in Eq.~(\ref{axion_mass}) we see that the axion is essentially massless down to a temperature $T = O({\rm GeV})$. This fact is extremely important when axions are studied in a cosmological context as in this paper. Defining the misalignment angle as
\begin{equation}
\theta(x) = \frac{a(x)}{f_a},
\end{equation}
the evolution of the zero mode of the dynamical field $\theta(x)$ in a flat Friedmann-Robertson-Walker metric is
\begin{equation}\label{eq_motion}
\ddot{\theta} + 3H(T)\,\dot{\theta}
+\frac{1}{f_a^2}\frac{\partial V(\theta)}{\partial\, \theta} = 0.
\end{equation}
A dot indicates a derivative with respect to time, $H(T)$ is
the Hubble parameter and the axion potential is
\begin{equation}
V(\theta) = m_a^2(T)f_a^2(1-\cos\theta).
\end{equation}
For small $\theta$, the potential is approximately harmonic, $V(\theta)\approx\frac{1}{2}m^2_a(T)f^2_a\theta^{2}$. In this case
the equation of motion Eq.~(\ref{eq_motion}) becomes
\begin{equation}\label{eq_motion1}
\ddot{\theta} + 3H(T)\,\dot{\theta} + m^2_a(T)\theta = 0.
\end{equation}
When $T \gg \Lambda$, the axion is massless, see Eq.~(\ref{axion_mass}), and a solution to Eq.~(\ref{eq_motion1}) is $\theta \equiv \theta_i = {\rm const}$, where $\theta_i$ is the initial value of the misalignment field, the so-called initial misalignment angle. The axion field is frozen at the value of $\theta_i$ from the onset of production until the axion mass becomes relevant and the axion field begins to oscillate. This occurs below a temperature $T_1 = O({\rm GeV})$ defined by
\begin{equation} \label{def_T1}
3H(T_1) = m_a(T_1).
\end{equation}
To find $T_1$ in terms of $m_a$, we use the Friedmann equation for a radiation-dominated Universe,
\begin{equation} \label{hubble_standard}
H(T) = \bigg[\frac{8\pi^3g_*(T)}{90M_{Pl}^2}\bigg]^{1/2}T^2 \approx 1.66\sqrt{g_*(T)}\frac{T^2}{M_{Pl}},
\end{equation}
where $g_*(T)$ is the total number of effective degrees of freedom at temperature $T$ \cite{kolb}. For the range of temperatures of interest in this paper, one has
\begin{equation}
g_*(T) =
\begin{cases}
61.75, & {\rm for}\,\, T \gtrsim \Lambda,\\
10.75, & {\rm for}\,\, \Lambda \gtrsim T \gtrsim 4{\rm MeV},\\
 3.36, & {\rm for}\,\, T \lesssim 4{\rm MeV}.
\end{cases}
\end{equation}
Inserting Eqs.~(\ref{axion_mass}) and~(\ref{hubble_standard}) in
Eq.~(\ref{def_T1}) gives
\begin{equation} \label{T_1}
T_1 =
\begin{cases}
\big(\frac{b\,m_a\,M_{Pl}\Lambda^4}{4.98\sqrt{g_*(T_1)}}\big)^{1/6} = 618{\rm MeV} \bigg(\frac{10^{12}{\rm GeV}}{f_a}\bigg)^{1/6}, & T \gtrsim \Lambda,\\
\big(\frac{m_a\,M_{Pl}}{4.98\sqrt{g_*(T_1)}}\big)^{1/2} = 68.1{\rm MeV} \bigg(\frac{10^{18}{\rm GeV}}{f_a}\bigg)^{1/2}, & T \lesssim \Lambda.\\
\end{cases}
\end{equation}
Since we expect $f_a \sim 10^{12}\,$GeV, the axion starts oscillating at $T_1 \sim 618\,$MeV. However, we will see that it is also possible for $f_a$ to be of order the GUT scale ($\sim 10^{16}\,$GeV) or even the Planck scale ($\sim 10^{19}\,$GeV) and coherent axion oscillations start later.

\section{WMAP Bounds on ${\bm H_I}$}

The Hubble expansion rate at the end of inflation $H_I$ can be constrained using data from WMAP5 plus BAO and SN. The curvature perturbation spectrum $\Delta^2_{\mathcal{R}}(k_0)$ at fixed wave number $k_0 = 0.002{\rm Mpc}^{-1}$ has been measured as \cite{komatsu}
\begin{equation}\label{measure_spectrum}
\Delta^2_{\mathcal{R}}(k_0) = (2.445 \pm 0.096)\times 10^{-9}.
\end{equation}
The tensor-to-scalar ratio $r$ has been constrained to be
\begin{equation} \label{measure_r}
r \equiv \frac{\Delta^2_h(k_0)}{\Delta^2_{\mathcal{R}}(k_0)} < 0.22 \quad \hbox{\rm at 95\% CL}.
\end{equation}
Eqs.~(\ref{measure_spectrum}) and~(\ref{measure_r}) can be combined to give an upper bound on the spectrum of primordial gravitational waves of about
\begin{equation}
\Delta^2_h(k_0) \lesssim 5.38\times 10^{-10}.
\end{equation}
Expressing $\Delta^2_h(k_0)$ in terms of $H_I$,
\begin{equation}
\Delta^2_h(k_0) = \frac{2 H^2_I}{\pi^2 M^2_{Pl}},
\end{equation}
leads to an upper bound on $H_I$,
\begin{equation} \label{HI_bound}
H_I < 6.29 \times 10^{14}{\rm ~GeV}.
\end{equation}
A lower limit on $H_I$ comes from requiring the Universe to be radiation-dominated at $T \simeq 4\,$MeV, so that primordial nucleosynthesis can take place \cite{hannestad}. Equating the highest temperature of the radiation
\begin{equation} \label{maximum_T}
T_{MAX} \sim (T^2_{RH} H_I M_{Pl})^{1/4},
\end{equation}
to the smallest allowed reheating temperature $T_{RH} = 4\,$MeV gives
\begin{equation}
H_I > H(T_{RH}) = 7.2 \times 10^{-24}{\rm ~GeV}.
\end{equation}

\section{Bounds on axion fluctuations}

If the axion energy scale is lower than the Gibbons-Hawking temperature $T_{GH} = H_I/2\pi$ \cite{birrell},
\begin{equation} \label{scenario_I}
f_a < T_{GH},
\end{equation}
the PQ symmetry is restored after inflation (Scenario I). When the Universe expands and cools down to a temperature $\sim f_a$, the PQ symmetry breaks again. Different values of the misalignment angle $\theta_i(x)$ are present within one Hubble volume \cite{kolb}, giving rise to fluctuations that are adiabatic as observed in the CMB spectrum.

On the contrary, non-adiabatic fluctuations are generated when (a) the PQ symmetry breaks before the end of inflation and (b) it is not restored afterwards (Scenario II). Condition (a) requires
\begin{equation} \label{scenario_II}
f_a > T_{GH}.
\end{equation}
Condition (b) requires
\begin{equation} \label{T_max}
f_a > T_{MAX},
\end{equation}
where $T_{MAX}$ is given in Eq.~(\ref{maximum_T}). Thus, non-adiabatic axion fluctuations arise when
\begin{equation} \label{(a)}
f_a > \max\{T_{GH},T_{MAX}\}.
\end{equation}

Theories where the inflaton decays into fermions favor $T_{RH} \lesssim 10^{12}\,$GeV \cite{linde}, which is consistent with assuming $T_{MAX} < T_{GH}$. In this scenario Eq.~(\ref{scenario_II}) suffices. The initial misalignment angle $\theta_i$ has a single value within one Hubble volume, since it was causally connected at the onset of production. The axion mass is negligible, see Eq.~(\ref{axion_mass}). Quantum fluctuations $\delta a(x)$ in the axion scalar field $a(x)$ have variance \cite{birrell}
\begin{equation} \label{fluctuations}
\langle|\delta a(x)|^2\rangle = \bigg(\frac{H_I}{2\pi}\bigg)^2.
\end{equation}
It follows that fluctuations in the misalignment angle field $\theta(x) = a(x)/f_a$ have variance
\begin{equation} \label{standard_deviation}
\sigma^2_{\theta} = \bigg(\frac{H_I}{2\pi f_a}\bigg)^2.
\end{equation}

Under condition Eq.~(\ref{(a)}) axion isocurvature perturbations are present during inflation and are constrained by WMAP5. Defining the power spectrum of axion perturbations $\Delta_a^2(k) = \langle|\delta \rho_a/\rho_a|^2\rangle$, one finds
\begin{equation} \label{axion_perturbations}
\Delta^2_a(k) = \frac{H^2_I}{\pi^2 \theta^2_i f^2_a}.
\end{equation}
The axion entropy-to-curvature perturbation ratio is then
\begin{equation}
\frac{\Delta^2_a(k_0)}{\Delta^2_{\mathcal{R}}(k_0)} =
\frac{H_I^2}{\pi^2 \Delta^2_{\mathcal{R}}(k_0) \theta_i^2 f^2_a} ,
\end{equation}
or, introducing the axion adiabaticity $\alpha_0(k_0)$,
\begin{equation}
\frac{\Delta^2_a(k_0)}{\Delta^2_{\mathcal{R}}(k_0)} = \frac{\alpha_0(k_0)}{1-\alpha_0(k_0)}.
\end{equation}
The adiabaticity $\alpha_0$ is constrained by WMAP 5-year data to be
\begin{equation} \label{alpha}
\alpha_0 < 0.072\quad\quad \hbox{ at 95\% CL}.
\end{equation}
Using the value for $\Delta^2_{\mathcal{R}}(k_0)$ in Eq.~(\ref{measure_spectrum}) this bound can be rephrased as
\begin{equation} \label{adiabaticity}
\frac{H_I}{\theta_i f_a} < 4.17 \times 10^{-5}.
\end{equation}

\section{Present axion energy density} \label{Density of axion in standard cosmology}

The major interest for axions in astrophysics is that it is possible for these particles to account alone for all of the observed CDM. If this is true, axions must be in highly non-thermal equilibrium and probably form a Bose-Einstein condensate \cite{kolb,
sikivie}. The leading mechanism for producing such an axion population is the misalignment production (see e.g.~\cite{hertzberg, kolb, sikivie, fox} and references therein). Another contribution that is important in Scenario I is axion production from string decay \cite{sikivie}.

According to the misalignment mechanism, the axion number density at temperature $T_1$ is given by \cite{kolb, sikivie}
\begin{equation} \label{number_nonstandard}
n_a^{mis}(T_1) = \frac{1}{2}m_a(T_1)f_a^2\langle\theta_i^2
f(\theta_i)\rangle\chi.
\end{equation}
Here the factor $\chi$ models the temperature dependence of the axion mass around $T_1$ and depends on the number of quark flavors $N_f$ that are relativistic at $T_1$ \cite{turner}. We take $\chi = 1.44$, consistent with $N_f =3$. The function $f(\theta_i)$ accounts for anharmonicity in the axion potential, i.e. for a solution to the full axion field Eq.~(\ref{eq_motion}) instead of Eq.~(\ref{eq_motion1}). The function $f(\theta_i)$ is
of order one for $|\theta_i| \lesssim 3$ and is logarithmical divergent for $\theta_i \to \pi$. It is discussed at the end of this Section.

The number density at present time is found by imposing the conservation
of the comoving axion number density in the form \cite{kolb}:
\begin{equation} \label{conservation}
\delta\!\left(\frac{n_a^{mis}(T)}{s(T)}\right) = 0,
\end{equation}
where $s(T)$ the entropy density. One has
\begin{equation} \label{entropy}
s(T) = \frac{2\pi^{2}}{45}g_{*S}(T)T^3.
\end{equation}
We use the approximation of the same number of degrees of freedom for the entropy and the total energy density for all $g_{*S}(T)$ with $T \gtrsim 4\,$MeV; the present value for $g_{*S}(T_0) = 3.91$ differs from $g_{*}(T_0) = 3.36$ \cite{kolb}.
Thanks to Eqs.~(\ref{number_nonstandard}),~(\ref{conservation})
and~(\ref{entropy}), the present axion energy density $\rho_a^{mis}(T_0) =
m_an_a^{mis}(T_0)$ is
\begin{equation} \label{energydensity}
\rho_a^{mis}(T_0) =
\frac{m_a\,m_a(T_1)s(T_0)}{2s(T_1)}f_a^2\langle\theta_i^2
f(\theta_i)\rangle\chi.
\end{equation}
Dividing the last equation by the critical density $\rho_c = 3H^2_0
M^2_{Pl}/8\pi$ and using Eq.~(\ref{axion_mass}) for $m_a(T_1)$, the
cosmologically relevant ratio $\Omega^{mis}_a = \rho_a/\rho_c$ is
\begin{equation}
\Omega^{mis}_a = \frac{Q^4s(T_0)}{2\rho_c s(T_1)}
\langle\theta_i^2 f(\theta_i)\rangle\chi \left\{
\begin{array}{c}
b \big(\frac{\Lambda}{T_1}\big)^4 ,\quad f_a\lesssim \hat{f}_a,\\
\quad 1, \quad\quad f_a\gtrsim \hat{f}_a.\\
\end{array}
\right.
\end{equation}
Or, if we insert the expression for $T_1$ previously computed in
Eq.~(\ref{T_1}) and substitute the numerical values $\chi = 1.44$, $b =
0.018$, we obtain
\begin{equation} \label{standarddensity}
\Omega^{mis}_a h^2 =
\begin{cases}
0.236\langle\theta_i^2 f(\theta_i)\rangle(\frac{f_a}{10^{12}{\rm GeV}})^{7/6}, & f_a \lesssim \hat{f}_a,\\
0.0051\langle\theta_i^2 f(\theta_i)\rangle(\frac{f_a}{10^{12}{\rm GeV}})^{3/2}, & f_a \gtrsim \hat{f}_a.\\
\end{cases}
\end{equation}
The scale $\hat{f}_a$ is computed by equating the two
expressions for $\Omega_a^{mis} h^2$ above:
\begin{equation}
\hat{f}_a = 9.91\times 10^{16}{\rm GeV}.
\end{equation}
The two cases in Eq.~(\ref{standarddensity}) reflect the dependence of the axion
mass $m_a(T)$ on the temperature $T$ in Eq.~(\ref{axion_mass}).

The factor $\langle\theta_i^2f(\theta_i)\rangle$ assumes different values in the two cases in which the axion is formed after inflation, $f_a < T_{GH}$ (Scenario I), or is present during inflation, $f_a >T_{GH}$ (Scenario II). In Scenario I, the variance is zero because there are no axion quantum fluctuations from inflation, but $\theta_i$ is not uniform over one Hubble volume, so $\theta_i^2$ is averaged over its possible values as
\begin{equation}
\langle\theta_i^2f(\theta_i)\rangle = \frac{1}{2\pi}\int_{-\pi}^{\pi}\theta_i^2 \,f(\theta_i)\, d\theta_i.
\end{equation}
If $f(\theta_i)$ is assumed to be 1, one finds the usual result $\langle\theta_i^2\rangle = \pi^2/3$. If $f(\theta_i)$ is taken to be the more realistic expression in Eq.~(\ref{anharmonicity_function}) below, one finds
\begin{equation} \label{theta_scenario_I}
\langle\theta_i^2f(\theta_i)\rangle = 8.77.
\end{equation}
Other expressions for $f(\theta_i)$ lead to similar values \cite{turner}.

In Scenario II, the initial misalignment field has a single value $\theta_i$ over one Hubble volume, so $\langle\theta_i\rangle = \theta_i$. The variance of the misalignment field is given by Eq.~(\ref{standard_deviation}), so
\begin{equation} \label{theta_scenario_II}
\langle\theta_i^2f(\theta_i)\rangle = \left( \theta_i^2 + \sigma_{\theta}^2 \right) \, f(\theta_i) = \left[ \theta_i^2 + \bigg(\frac{H_I}{2\pi f_a}\bigg)^2 \right] \, f(\theta_i).
\end{equation}
The initial misalignment angle $\theta_i$ in this scenario can assume any value between $-\pi$ and $\pi$. When $\theta_i \ll 1$, the axion energy scale $f_a \gg 10^{12}\,$GeV if the axion is to be the CDM particle \cite{linde1}.

We now discuss the anharmonicity factor $f(\theta_i)$. This factor accounts for the corrections that must be made to the solutions of the linear Eq.~(\ref{eq_motion1}) to obtain the solutions of the full non-linear Eq.~(\ref{eq_motion}). The anharmonicity factor $f(\theta_i)$ has the following limiting behavior:
\begin{equation} \label{condition}
\lim_{\theta_i \rightarrow\, 0} f(\theta_i) = 1\,\,; \quad\quad \lim_{\theta_i \rightarrow \pm \pi} f(\theta_i) = +\infty.
\end{equation}
The problem of finding a specific shape for $f(\theta_i)$ has been considered by many authors. Turner \cite{turner} integrates Eq.~(\ref{eq_motion}) numerically and describes how the anharmonicity factor can be computed, but does not give an explicit formula. Lyth \cite{lyth1} follows the idea in \cite{turner} and performs an explicit calculation, obtaining the behavior $f(\theta) \sim \ln^{1.175}(1-\theta_i/\pi)$ for $\theta_i > 0.9\pi$; he also comments that his result differs from Turner's by a factor of two. The exponent 1.175 comes from Lyth's dependence of $m_a(T)$ on $T$, which is $m_a(T) \sim T^{-3.7}$, different from ours. Finally, Strobl and Weiler \cite{strobl} and Bae {\it et al.} \cite{bae} perform a more precise numerical analysis following \cite{turner}, and confirm the result in \cite{lyth1} for the behavior of $f(\theta_i)$ around $\theta_i = \pi$.

In this paper, we adopt the following analytic expression for $f(\theta_i)$, which extends Lyth's formula to values of $\theta_i < 0.9\pi$ and symmetrically to negative values of $\theta_i$:
\begin{equation} \label{anharmonicity_function}
f(\theta_i) = \left[\ln\left(\frac{e}{1-\theta_i^2/\pi^2}\right)\right]^{7/6}.
\end{equation}
This expression has the limiting behaviors in Eq.~(\ref{condition}) and is analytic in the range $(-\pi, \pi)$, so the computation of the level curves in Fig.~\ref{classical} below can be carried out analytically. The different exponent in the power of the logarithm, $7/6 \sim 1.167$ instead of Lyth's $1.175$, comes from the different dependence of the axion mass $m_a(T)$ on $T$ in Eq.~(\ref{axion_mass}), $m_a(T) \sim T^{-3.7}$ in Lyth's paper and $m_a(T) \sim T^{-4}$ in this paper.

Other authors find different expressions for the axion energy density.

Hertzberg {\it et al.} \cite{hertzberg} use different values for $\chi$, which range from $\chi = 1$ (which they call a ``moderate'' value) to $\chi = 1/20$ (which they call a ``conservative'' value). However, the authors in \cite{hertzberg} do not account for anharmonicities in the axion potential near $\theta_i = \pi$. As shown in Section~\ref{sec:results}, the anharmonicity factor is essential to obtain the correct behavior of the isocurvature fluctuation bound at relatively small values of $f_a$, because the initial misalignment angles that give the correct axion density are not very small. We choose $\chi = 1.44$, but most importantly we differ from \cite{hertzberg} in that we account for the behavior of the anharmonicity function $f(\theta_i)$. With this prescription, it turns out that the figures shown in \cite{hertzberg} are modified when the function $f(\theta_i)$ ceases to be of order one. This happens for $f_a \lesssim 10^{11}\,$GeV, as shown in Fig.\ref{classical} below.

Sikivie \cite{sikivie} studies the axion energy density in the case $T_1 > \Lambda$, and finds $\Omega_a = 0.15\,\theta_i^2\,(f_a/10^{12}{\rm GeV})^{7/6}$. His expression differs from ours in Eq.~(\ref{standarddensity}) because of the different numerical factors used in \cite{sikivie}, namely $\chi = 1$, $b=5/12$ and $T_1 = 1{\rm GeV}(10^{12}{\rm GeV}/f_a)^{1/6}$.

\section{Results}
\label{sec:results}

In this section we show the region of axion parameter space in which the axion is 100$\%$ of the cold dark matter. In other words, we assume
\begin{equation} \label{CDM}
\Omega_a h^2 = \Omega_{CDM} h^2 = 0.1131 \pm 0.0034 \quad {\rm at}\, 68\% \,{\rm CL}.
\end{equation}
Here $\Omega_{CDM}$ is the density of cold dark matter in units of the critical density and $h$ is the Hubble constant in units of 100\,km s$^{-1}$Mpc$^{-1}$.

\begin{figure}[t]
\includegraphics[width=13cm]{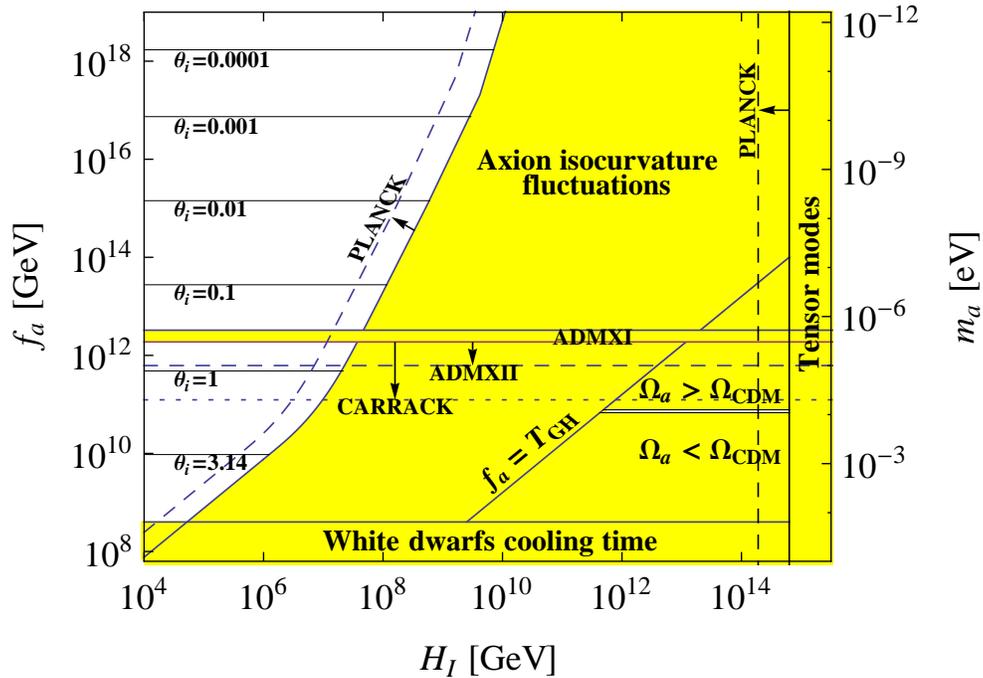}
\caption{Region of axion parameter space where the axion is 100\% of the cold dark matter. The axion mass scale on the right corresponds to Eq.~(\ref{eq:axionmass}) with $U(1)_{PQ}$ color anomaly $N=1$. When the PQ symmetry breaks after inflation ($f_a < H_I/2\pi$), the axion is the CDM particle if $f_a =  (7.27 \pm 0.25)\times 10^{10}{\rm ~GeV}$, or $m_a=(85\pm3){\rm ~\mu eV}$, which is the narrow horizontal window shown on the right (we plot a 3$\sigma$ window to make it visible). If the axion is present during inflation ($f_a > H_I/2\pi$), axion isocurvature perturbations constrain the parameter space to the region on the top left, which is marked by the values of $\theta_i$ necessary to obtain 100\% of the CDM density. Other bounds indicated in the figure come from astrophysical observations of white dwarfs cooling times and the non-observation of tensor modes in the Cosmic Microwave Background fluctuations. Dashed lines and arrows indicate the future reach of the PLANCK satellite and the ADMX and CARRACK microwave cavity searches.}
\label{classical}
\end{figure}

The parameter space is labeled by the axion energy scale $f_a$, the initial misalignment angle $\theta_i$, and the Hubble parameter during inflation $H_I$. Results are shown in Fig.~\ref{classical}.

The region labeled ``Tensor modes'' shows the constraint on $H_I$ in Eq.~(\ref{HI_bound}) coming from the WMAP5 plus BAO and SN observations \cite{komatsu}. The newly launched PLANCK satellite will improve the actual measurement on the tensor-to-scalar ratio $r$ by at least one order of magnitude \cite{planck}. If PLANCK will not detect gravitational waves and $r < 0.02$, $H_I$ will be bound by $H_I \lesssim 2\times 10^{14}\,$GeV. This forecast measurement is shown by a vertical dashed line labeled ``PLANCK''.

The region labeled ``White dwarfs cooling time'' is excluded because there one would have an excessively small cooling time in white dwarfs. These methods set a limit on the axion energy scale of \cite{raffelt}
\begin{equation}
f_a > 4 \times 10^{8} {\rm GeV}.
\end{equation}
Assuming $N=1$ in Eq.~(\ref{eq:axionmass}), this corresponds to $m_a < 15$ meV.

The line
\begin{equation}
f_a = T_{GH} = H_I/2\pi
\end{equation}
divides the region where the PQ symmetry breaks after inflation, $f_a < T_{GH}$ (Scenario I), from the region where the axion field is present during inflation, $f_a > T_{GH}$ (Scenario II).

The region marked as ADMXI has been excluded by a direct search of axions CDM in a Sikivie microwave cavity detector \cite{microwave_sikivie} by the ADMX Phase I experiment \cite{asztalos, duffy}. The window shown corresponds to $1.9\,{\rm \mu eV} < m_a < 3.3\,{\rm \mu eV}$, valid for the KSVZ axion model. A narrower DFSV axionic window, not shown in the figure, has also been ruled out. The dashed line labeled ``ADMXII'' shows the forecast axionic region to be probed in the ADMX Phase II, which would search for axions with mass up to $10{\rm \mu eV}$, or $f_a = 6.2\times 10^{11}\,$GeV. The proposed CARRACK II experiment is a cavity search that will look for axions with mass up to $50{\rm \mu eV}$ \cite{carrack}.

For Scenario I, since $\langle\theta_i^2 f(\theta_i)\rangle$ does not depend on $H_I$ or $\theta_i$, the value of $\Omega_a^{mis}$ depends on $f_a$ only. From Eq.~(\ref{standarddensity}) with $f_a<\hat{f}_a$ and Eq.~(\ref{theta_scenario_I}), we find
\begin{equation} \label{mis}
\Omega^{mis}_a h^2 = 2.07 \left(\frac{f_a}{10^{12}{\rm GeV}}\right)^{7/6}.
\end{equation}
As pointed out in \cite{sikivie}, axionic string decays are an important mechanism of axion production in Scenario I (in Scenario II all string defects are washed out by inflation). The present density of axions produced in string decays is \cite{sikivie}
\begin{equation} \label{string}
\Omega^{str}_a h^2 = 0.34 \left(\frac{f_a}{10^{12}{\rm GeV}}\right)^{7/6}.
\end{equation}
The total axion energy density is therefore
\begin{equation}
\Omega_a h^2 = \Omega^{mis}_a h^2 + \Omega^{str}_a h^2 = 2.41 \left(\frac{f_a}{10^{12}{\rm GeV}}\right)^{7/6}.
\end{equation}
The value of $f_a$ such that the axion is 100\% of the cold dark matter, $\Omega_a h^2 = \Omega_{CDM} h^2$, is then
\begin{equation} \label{PQ_strength}
f_a = (7.27 \pm 0.25)\times 10^{10}{\rm GeV}.
\end{equation}
This band is drawn in Fig.~\ref{classical} as the horizontal window on the lower right. The result in Eq.~(\ref{PQ_strength}) is one order of magnitude lower than the usually quoted value for the PQ energy scale, $f_a \approx 10^{12}\,$GeV. This difference by one order of magnitude comes from equating $\Omega_a$ with the precise result in Eq.~(\ref{CDM}) for $\Omega_{CDM}$, that is $\Omega_a\sim 0.1$, whereas seminal works in axion cosmology computed $f_a$ from the equation $\Omega_a = 1$. We also include in $\Omega_a$ the contribution from axionic strings decay, Eq.~(\ref{string}), and the careful derivation of Eq.~(\ref{mis}).

Assuming a $U(1)_{PQ}$ color anomaly $N=1$, the $\Omega_a=\Omega_{CDM}$ band corresponds to an axion mass
\begin{equation}
m_a = (85 \pm 3) {\rm ~\mu eV}.
\end{equation}

In Scenario II, axion isocurvature fluctuations are present and lead to the bound in Eq.~(\ref{adiabaticity}). Together with the expression for $\Omega_a^{mis}$ in Eq.~(\ref{standarddensity}) and the condition of 100\% CDM in Eq.~(\ref{CDM}), the adiabaticity bound excludes the shaded region in the center of Fig~\ref{classical}. The PLANCK satellite is expected to improve the current bounds on the axion isocurvature fluctuations by at least one order of magnitude \cite{planck}. The dashed line on the left of Fig.~\ref{classical} shows the new bound on the allowed region when Eq.~(\ref{alpha}) is replaced by $\alpha_0 < 7\times 10^{-3}$.

The leftmost boundary of this region contains two kinks and can be approximated with
\begin{equation} \label{leftmostboundary}
f_a = \begin{cases}
7.64\times 10^{9}{\rm ~GeV} \left(\frac{H_I}{10^6{\rm ~GeV}}\right) , & H_I < 9.96 \times 10^6 {\rm ~GeV} , \\
1.97\times 10^{13}{\rm ~GeV} \left(\frac{H_I}{10^8{\rm ~GeV}} \right)^{12/5} , & 9.96 \times 10^6 {\rm ~GeV} < H_I < 3.43 \times 10^9 {\rm ~GeV}, \\
6.74 \times 10^{18}{\rm ~GeV} \left( \frac{H_I}{10^{10}{\rm ~GeV}} \right)^4 , & H_I > 3.43 \times 10^9 {\rm ~GeV}.
\end{cases}
\end{equation} The upper kink occurs at $f_a = \hat{f}_a$ and is due to the change in the dependence of the axion mass on the temperature,  Eq.~(\ref{axion_mass}). The lower and smoother kink around $f_a \sim 10^{11}$GeV arises from the fact that the anharmonicity function $f(\theta_i)$ differs from one at values of $f_a$ smaller than $10^{11}$ GeV, see Eq.~(\ref{anharmonicity_function}). Notice that the simple proportionality $f_a \propto H_I$ at small $H_I \lesssim 10^7$ GeV is independent of the detailed form assumed for the function $f(\theta_i)$ near $\theta_i=\pi$, and derives in a straightforward way from Eq.~(\ref{adiabaticity}).

\begin{figure}[t]
\includegraphics{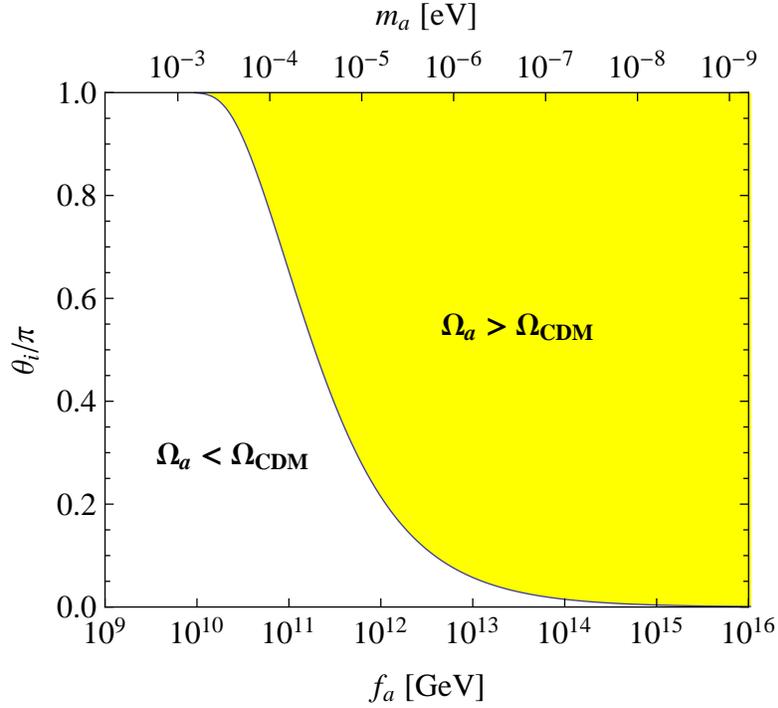}
\caption{The misalignment angle $\theta_i$ necessary for the axion to be 100\% of the cold dark matter in Scenario II ($f_a>H_I/2\pi$), as a function of $f_a$. Above the curve, $\Omega_a>\Omega_{CDM}$. For $f_a \gtrsim 10^{17}$ GeV, one has $\theta_i \simeq 0.001 (f_a/10^{17}\,{\rm GeV})^{-3/4}$; in particular, for $f_a \gtrsim 10^{19}$ GeV, the initial misalignment angle $\theta_i$ has to assume values $\theta_i \lesssim 10^{-5}$.}
\label{theta_plot}
\end{figure}

In the remaining region on the left of Fig.~\ref{classical}, the axion can be 100$\%$ of the cold dark matter, provided the value of the initial misalignment angle $\theta_i$ is chosen appropriately. The $\theta_i$ contours in the figure indicate the appropriate values of $\theta_i$ for a given $H_I$ and $f_a$. Notice that the $\theta_i$ contours are horizontal, i.e.\ independent of $H_I$, since in that region the contribution from $\sigma_{\theta}^2$ in Eq.~(\ref{theta_scenario_II}) is negligible compared to $\theta_i^2$. This allows us to show the full relation between $f_a$ and $\theta_i$ imposed by the constraint $\Omega_a=\Omega_{CDM}$. We find
\begin{equation} \label{const_theta}
\frac{f_a}{10^{12}\,{\rm GeV}} = \begin{cases}
\big(\frac{\Omega_{CDM}\,h^2}{0.236\,\theta_i^2\, f(\theta_i)}\big)^{6/7}, & f_a \lesssim \hat{f}_a {\rm ~or~} \theta_i \gtrsim 0.001,\\
\big(\frac{\Omega_{CDM}\,h^2}{0.0051\,\theta_i^2\, f(\theta_i)}\big)^{2/3}, & f_a \gtrsim \hat{f}_a {\rm ~or~} \theta_i \lesssim 0.001.
\end{cases}
\end{equation}
This is illustrated in Fig.~\ref{theta_plot} for $\Omega_{CDM} h^2$ in Eq.~(\ref{CDM}).

For $f_a \gtrsim 10^{17}$ GeV ($m_a \lesssim 10^{-10}{\rm~eV}$ or $\theta_i \lesssim 0.001$), one has $f(\theta_i)\simeq 1$ and Eq.~(\ref{const_theta}) simplifies to
\begin{equation}
\theta_i \simeq 0.84\times 10^{-3} \left( \frac{f_a}{10^{17}\,{\rm GeV}} \right)^{-3/4}  , \qquad {\rm for~} f_a\gtrsim10^{17}{\rm ~GeV},
\end{equation}
or
\begin{equation}
\theta_i \simeq 1.2\times 10^{-3} \left(\frac{m_a}{10^{-10}{\rm ~eV}}\right)^{3/4} , \qquad {\rm for~} m_a\lesssim 10^{-10}{\rm ~eV}.
\end{equation}
In particular, for $f_a \gtrsim 10^{19}$ GeV, the initial misalignment angle $\theta_i$ has to assume values $\theta_i \lesssim 10^{-5}$. This was also noted in \cite{fox, bae}. These small values of $\theta_i$ may be uncomfortable in a cosmological scenario.

In the other limit of $\theta_i \simeq \pi$, the form of the function $f(\theta_i)$ assumed in Eq.~(\ref{anharmonicity_function}) gives
\begin{equation}
\pi-\theta_i \simeq \frac{e\pi}{2} e^{-C/f_a}, \qquad {\rm for~}f_a\lesssim2\times10^{10}{\rm ~GeV},
\end{equation}
with $C=7.48\times10^{10}$ GeV. So, as $\theta_i$ approaches $\pi$ from below, the corresponding $f_a$ approaches 0. This gives rise to the linear dependence of $f_a$ on $H_I$ in the lower left corner of Fig.~\ref{classical}.

\section{Discussion}

We have seen that depending on the ratio $f_a/H_I$, two scenarios are possible for the axions to be 100\% of the cold dark matter, which we called Scenario I ($f_a<H_I/2\pi$) and Scenario II ($f_a>H_I/2\pi$). In Scenario I, the Hubble scale $H_I$ is bounded from below to $H_I \gtrsim 10^{11}$GeV (leftmost edge of the $\Omega_a=\Omega_{CDM}$ horizontal window in Fig.~\ref{classical}). This window can disappear completely if the limit from tensor modes moves to the left beyond $H_I \gtrsim 10^{11}$GeV. If the limit on $r$ would become more stringent than $r\gtrsim 10^{-8}$, Scenario I would have to be abandoned in favor of Scenario II.

We remark that for very large values of $f_a$ in Scenario II, the initial misalignment angle $\theta_i$ has to be chosen very small for the axion to be 100\% of the CDM, see Eq.~(\ref{const_theta}) and Fig.~\ref{theta_plot}. This may undermine the axion field as a dynamical solution to the strong CP problem, in that it would have to be fixed to a small value as an initial condition.

Grand Unification Theory (GUT) models that contain axions predict that $f_a$ should be of the order of the GUT scale, $\sim 10^{16}$GeV \cite{witten}. In a variety of string theory models \cite{kim, witten}, the PQ energy scale results in the range $10^{16}{\rm GeV} < f_a < 10^{18}{\rm GeV}$. From Fig.~\ref{classical} we see that this range of $f_a$ values cannot be reconciled with axions as 100\% CDM in Scenario I, while they can be in Scenario II provided $H_I\lesssim 10^{9}$GeV.

We studied the possibility for the axion to form the totality of the cold dark matter in the light of WMAP5 observations. There are two scenarios for this to happen. In Scenario I, in which the PQ symmetry is restored after inflation ($f_a<H_I/2\pi$), one needs $H_I>4.57\times10^{11}$ GeV and $f_a=(7.27\pm0.25)\times10^{10}\,$GeV, corresponding to an axion mass $m_a= (85\pm3){\rm~\mu eV}$. In Scenario II, in which the PQ symmetry breaks during inflation and is not restored afterwards ($f_a>H_I/2\pi$), one can have $f_a>4\times10^8$GeV ($m_a<15$ meV) and $H_I$ constrained as in Fig.~\ref{classical} (see Eq.~(\ref{leftmostboundary})). Moreover, in this region, the misalignment angle $\theta_i$ must be chosen appropriately, as Eq.~(\ref{const_theta}) and Fig.~\ref{theta_plot}. Large values of the PQ symmetric breaking scale $f_a$ then correspond to small values of the initial misalignment angle $\theta_i$. Determining the PQ energy scale $f_a$, or the axion mass $m_a$, is thus tightly related to constraining inflation parameters and future extensions of the Standard Model.

\begin{acknowledgements}
The authors would like to thank P. Sikivie for helpful suggestions and discussions. This work was partially supported by NSF award PHY-0456825.
\end{acknowledgements}

\end{document}